# Special relativity from a single scenario using the same strategy

Bernhard Rothenstein[1)] and Stefan Popescu[2)]
1) Politehnica University of Timisoara, Physics Department, Timisoara, Romania
2) Siemens AG, Erlangen, Germany

**Abstract.** *Following an approach proposed by Rosser[4] for deriving the transformation equations of volume charge density and current density we derive the transformation equations for the space-time coordinates of the same event, for the mass and the momentum of the same particle and for the electric and the magnetic field generated by the same distribution of electric charges.*

## 1. Introduction

Special relativity involves two inertial reference frames K(XOY) and K'(X'O'Y') in relative motion. The corresponding axes of the two frames are parallel to each other and the OX(O'X') axes are common. At the origin of time (*t=t'=0*) the origins of the two frames O(O') are located at the same point in space. K' moves with constant speed *V* relative to K in the positive direction of the common axes. Consider a physical system whose physical properties are measured by observers of the two frames. Any physical property is characterized by a physical quantity expressed as a product of a numerical value and a physical unit. In the general case when measuring a given quantity the observers of the two frames obtain different results, say $\Phi$ in K and $\Phi'$ in K'. The transformation equations establish a relationship between $\Phi$ and $\Phi'$. We distinguish between the two possible situations:

a. The physical system under consideration is at rest relative to K'. In this case, the physical quantities measured by observers from K' are proper physical quantities, whereas those measured by observers from K are non-proper physical quantities. The transformation equation relates a proper physical quantity $\Phi'$ measured in K' to a non-proper one $\Phi$ measured in K.
b. The physical system moves relative to both K and to K'. In this case the transformation equation relates two non-proper physical quantities $\Phi$ and $\Phi'$ (Lorentz-Einstein transformations).

The transformation equations of the type (a) are derived by many authors as particular cases of the transformation equations of the type (b).

The postulates of special relativity and thought experiments solve the problem in the case (a), resulting in the transformation equation for relativistic velocities.[1,2,3] The purpose of our paper is to show that if the (a) problem is solved then the (b) problem can be solved following the same strategy, in all chapters of physics: kinematics, dynamics and electrodynamics. Doing so we follow a strategy proposed by Rosser[4] for deriving the transformation equations for volume charge density and current density.

The scenario we propose involves the inertial reference frames K(XOY), K'(X'O'Y') and $K^0(X^0O^0Y^0)$. $K^0$ is the inertial reference frame in which the physical system being studied is in a state of rest and observers of this frame measure proper physical quantities. The frame $K^0$ moves with speed $u_x$ relative to K and with speed $u'_x$ relative to K', whereas K' moves with speed $V$ relative to K, all in the positive direction of the common axes. As shown in Figure 1 we find at rest at the origin $O^0$ of $K^0$ a particle of proper mass $m_0$ and a clock $C^0$. Also at rest along the $O^0X^0$ axis we find a uniform distribution of electric charges characterized by a linear density $\lambda_0$.

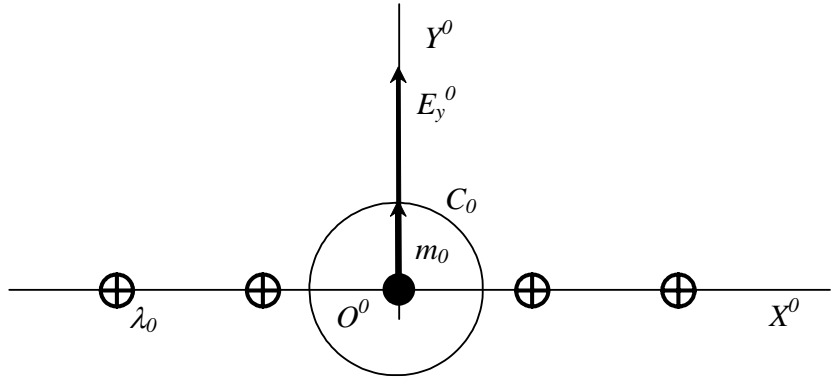

**Figure 1**. The scenario in its rest frame $K^0(X^0O^0Y^0)$. It involves a clock $C^0$, a particle of rest mass $m_0$ and a uniform distribution of electric charges located along the $X^0$ axis.

### 2. Relativistic kinematics

Consider the situation presented in Figure 2. Figure 2a shows the relative position of $K^0$ relative to K as detected from K at a given time $t$ displayed by the synchronized clocks of K. Figure 2b shows the relative position of $K^0$ relative to K' at a given time $t'$ displayed by the synchronized clocks of K'.

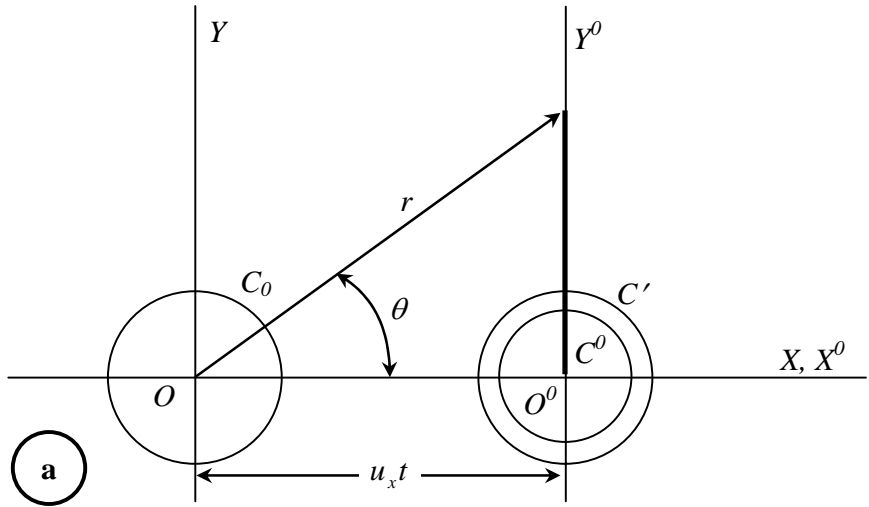

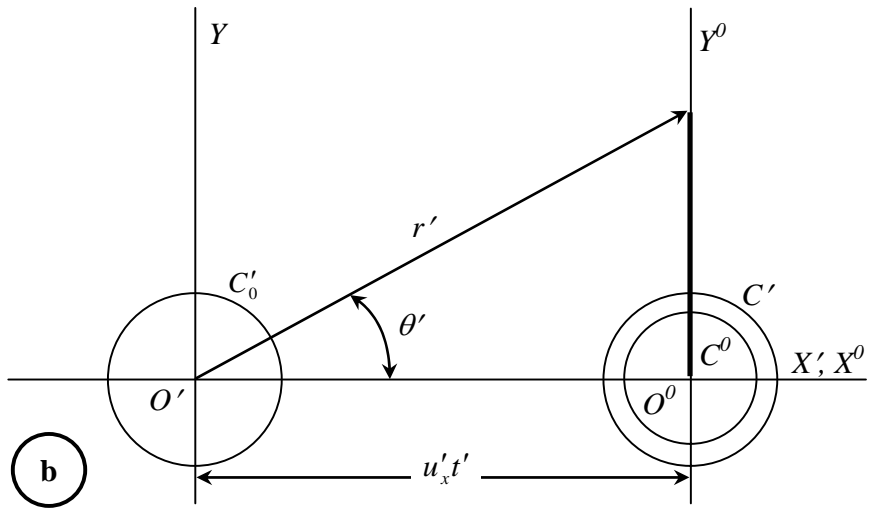

**Figure 2a**. The relative position of the reference frames K and $K^0$ as detected from the K frame at a given time t. The clock $C^0$ is located in front of a clock C of the K frame.
**Figure 2b**. The relative position of the reference frames K' and $K^0$ as detected from the K' frame. The clock $C^0$ is located in front of a clock C' of the K' frame/

Clock $C^0$, defined above, moves with velocity $u_x$ relative to K and with velocity $u'_x$ relative to K'. Without using the Lorentz-Einstein transformations we can show that[1,2,3]

$$u_x = \frac{u'_x + V}{1 + u'_x V/c^2} \tag{1}$$

and that

$$u'_x = \frac{u_x - V}{1 - u_x V / c^2}. \tag{2}$$

As we show in Figure 2a the clock $C^0$ is reading $t^0$ when it is located in front of a clock $C$ of K that is reading $t$. Without using the Lorentz-Einstein transformations we can show that the two clock readings are related by[2]

$$t = \frac{t^0}{\sqrt{1 - u_x^2 / c^2}}.$$

In the situation presented in Figure 2b the clock $C^0$ is reading $t^0$ when it is located in front of the clock $C'$ of K' that is reading $t$ and for the same reasons as above we have

$$t' = \frac{t^0}{\sqrt{1 - u'^2_x / c^2}}. \tag{3}$$

Eliminating $t^0$ between (2) and (3) and taking into account (1) and (2), we obtain that the readings of clocks $C$ and $C'$ are related by

$$t = t' \frac{1 + V u'_x / c^2}{\sqrt{1 - V^2 / c^2}} = \frac{t' + V x' / c^2}{\sqrt{1 - V^2 / c^2}} \tag{4}$$

or by

$$t' = t \frac{1 - V u_x / c^2}{\sqrt{1 - V^2 / c^2}} = \frac{t - V x / c^2}{\sqrt{1 - V^2 / c^2}} \tag{5}$$

where $x = u_x t$ (6) defines the position of clock $C^0$ in the frame K, whereas $x' = u'_x t'$ (7) defines its position relative to frame K'. Starting with (6) and expressing its right side as a function of physical quantities measured in the frame K', we obtain

$$x = u_x t = \frac{x' + V t'}{\sqrt{1 - V^2 / c^2}} \tag{6}$$

whereas starting with (7) in the same way we obtain

$$x' = u'_x t' = \frac{x - V t}{\sqrt{1 - V^2 / c^2}}. \tag{7}$$

The event associated with the fact that clocks $C^0$, $C$ and $C'$ defined above are located at the same point in space is characterized in K by its space-time coordinates $E(x = u_x t, y = 0, t)$ and in K' by its space-time coordinates $E'(x' = u't', y' = 0, t')$. By definition under such conditions $E$ and $E'$ represent the same event and the Lorentz-Einstein transformations (4), (6) and (5), (7) perform their transformation between K → K' and respectively K' → K.[5,6,7]

### 3. Relativistic dynamics

We consider the point-like particle located at rest at the origin $O^0$ of the frame $K^0$. Observers of this frame measure its proper (rest) mass $m_0$. Observers from K measure its relativistic mass $m$ (we will show later how this concept can be avoided), whereas observers from K' measure its relativistic mass $m'$. A thought experiment that involves the relativistic postulates and (1) and (2) shows that[8]

$$m = \frac{m_0}{\sqrt{1 - u_x^2 / c^2}} \tag{8}$$

and that

$$m' = \frac{m_0}{\sqrt{1 - u_x'^2 / c^2}}. \tag{9}$$

Eliminating $m_0$ between (8) and (9) and taking into account (1) and (2) we obtain that $m$ and $m'$ are related by

$$m = m' \frac{1 + V u_x' / c^2}{\sqrt{1 - V^2 / c^2}} = \frac{m' + V m' u_x'}{\sqrt{1 - V^2 / c^2}} \tag{10}$$

and

$$m' = m \frac{1 - V u_x / c^2}{\sqrt{1 - V^2 / c^2}} = \frac{m - V m u_x / c^2}{\sqrt{1 - V^2 / c^2}}. \tag{11}$$

On the right side of (10) and (11) we observe the presence of the terms $p_x = m u_x$ and $p_x' = m' u_x'$, which have the physical dimensions of momentum, representing the OX(O'X') components of the particle being considered, as measured by observers from K and K', respectively. They transform as

$$p_x = m u_x = \frac{p_x' + V m'}{\sqrt{1 - V^2 / c^2}} \tag{12}$$

and

$$p_x' = m' u_x' = \frac{p_x - V m}{\sqrt{1 - V^2 / c^2}}. \tag{13}$$

Multiplying both sides of (10) and (11) by $c^2$ and introducing the concept of relativistic energy $E = mc^2$ in K and $E' = m'c^2$ in K' these equations become

$$E = \frac{E' + V p_x'}{\sqrt{1 - V^2 / c^2}} \tag{14}$$

and

$$E' = \frac{E - V p_x}{\sqrt{1 - V^2 / c^2}}. \tag{15}$$

Expressed as a function of energy, (12) and (13) become

$$p_x = \frac{p'_x + VE'/c^2}{\sqrt{1-V^2/c^2}} \qquad (16)$$

and

$$p'_x = \frac{p_x - VE/c^2}{\sqrt{1-V^2/c^2}}. \qquad (17)$$

Equations (14)-(17) enable us to avoid the concept of relativistic mass which many authors prefer to avoid.[9]

**4. Aberration of light**

Thought experiments that do not involve light signals prove the invariance of distances measured perpendicular to the direction of relative motion.[10] Consider that the source of light $S_0$ located at the origin $O^0$ of $K^0$ emits a light signal at the origin of time, in the positive direction of the $O^0Y^0$ axis. (Figure 2a). After a given time of propagation the light signal generates the event $E^0(0, d=ct^0, t^0)$. The same event detected from the K frame (Figure 2b) has the space-time coordinates $E(0, r\sin\theta, t=r/c)$. Pythagoras' theorem tells us that

$$\cos\theta = \frac{u_x}{c}. \qquad (18)$$

Detecting the same experiment from K', we obtain

$$\cos\theta' = \frac{u'_x}{c}. \qquad (19)$$

Combining (18) and (19) we obtain that the angles $\theta$ and $\theta'$ are related by

$$\cos\theta = \frac{u_x}{u'_x}\cos\theta' = \frac{1+V/u'_x}{1+Vu'_x/c^2} = \frac{\cos\theta' + V/c}{1+\frac{V}{c}\cos\theta'} \qquad (20)$$

and

$$\cos\theta' = \frac{u'_x}{u_x}\cos\theta = \frac{\cos\theta - V/c}{1-\frac{V}{c}\cos\theta}. \qquad (21)$$

Simple trigonometry leads to the transformation equations for the other trigonometric functions of the angles involved.

**4. Lorentz-Einstein transformations for the polar and time coordinates of the same event**

Consider that a light signal starts to propagate at $t=t'=0$ from the point where the origins O and O' are located at that very moment. Detected from K, the light signal propagates along a direction $\theta$, whereas detected from K' it propagates along a direction $\theta'$ both in the positive direction of the common axes. Detected from K, the light signal generates the event $E(x = r\cos\theta, y = r\sin\theta, t = r/c)$, whereas detected from K' it generates the event $E'(x' = r'\cos\theta', y' = r'\sin\theta', t' = r'/c)$. The invariance of distances measured perpendicular to the direction of relative motion requires that

$$r\sin\theta = r'\sin\theta' \qquad (22)$$

from where we obtain that the length of the position vector of the point where the event takes place transforms as

$$r = r'\frac{\sin\theta'}{\sin\theta} = r'\frac{1 + \frac{V}{c}\cos\theta'}{\sqrt{1 - V^2/c^2}} = \frac{r' + \frac{V}{c}x'}{\sqrt{1 - V^2/c^2}}. \qquad (23)$$

The time coordinates transform as

$$t = \frac{r}{c} = t'\frac{1 + \frac{V}{c}\cos\theta'}{\sqrt{1 - V^2/c^2}} = \frac{t' + \frac{V}{c^2}x'}{\sqrt{1 - V^2/c^2}} . \qquad (24)$$

and the polar angles transform in accordance with (20).

The experience gained so far teaches us that we can obtain the inverse transformations by simply interchanging the corresponding un-primed physical quantities with primed ones and changing the sign of the relative velocity *V*.

As an exercise for the reader, we propose the derivation of the Lorentz-Einstein transformations for the space time coordinates of events $E(x = r\cos\theta, y = r\sin\theta, t = r/u)$ as detected from the frame K and $E'(x' = r'\cos\theta', y' = r'\sin\theta', t' = r'/u')$ as detected from K', generated by a tardyon that goes through the point where the origins O and O' are located at *t=t'=0* with velocities *u* and *u'* along directions which make the angles $\theta$ and $\theta'$ with the positive direction of the common axes.

### 5. Relativistic electrodynamics

Consider that the linear electric charge distribution presented in Figure 1 is at rest in $K^0$. Observers from this frame characterize it by a proper linear density $\lambda_0$. At a given point $M^0(0,d)$ of the $O^0Y^0$ axis it generates an electric field that has only a normal (proper) component given by[11]

$$E_y^0 = \frac{\lambda_0}{\varepsilon_0 d}. \qquad (25)$$

Taking into account the length contraction effect that affects the linear density of charge, the same electric field detected from K is

$$E_y = \frac{E_y^0}{\sqrt{1 - u_x^2/c^2}} \qquad (26)$$

whereas detected from K' it is

$$E'_y = \frac{E_y^0}{\sqrt{1 - u'^2_x/c^2}}. \qquad (27)$$

Combining (26) and (27), taking into account (1) and (2), we obtain

$$E_y = E'_y \frac{1 + Vu'_x/c^2}{\sqrt{1 - V^2/c^2}} = \frac{E'_y + VE'_y u'_x/c^2}{\sqrt{1 - V^2/c^2}}. \qquad (28)$$

On the right side of (28) we detect the presence of the term $E'_y u'_x/c^2$ (29) that reads $E_y u_x/c^2$ (30) in K and transforms as

$$\frac{E_y u_x}{c^2} = \frac{E'_y u'_x}{c^2} \frac{1 + V/u'_x}{\sqrt{1 - V^2/c^2}}. \qquad (31)$$

We can think of (28) and (29) as the OZ (O'Z') components of the vector product $(\mathbf{u}_x \times \mathbf{E}_y)/c^2$ and $(\mathbf{u}'_x \times \mathbf{E}'_y)/c^2$. Using for them the notations $B_z$ and respectively $B'_z$ (31) becomes

$$B_z = \frac{B'_z + VE'_y/c^2}{\sqrt{1 - V^2/c^2}}. \qquad (32)$$

We call $B_z$ and $B'_z$ the OZ(O'Z') components of the magnetic vector in K and in K', respectively.

### 6. Conclusions

Generalizing the results obtained above, we can say that a equation of Lorentz-Einstein transformation relates a scalar physical quantity for which we can define a proper value to a vector physical quantity which is the product of the scalar physical quantity and the velocity at which the physical system characterised by it moves relative to the reference frame being considered. With this the way to the concept of four vector is open.

### References

[1] W.N.Matews Jr. "Relativistic velocity and acceleration transformations from thought experiments." Am.J.Phys. **73**, 45-51 (2005) and references therein

[2] Margaret Stautberg Greenwood, "Relativistic addition of velocities using Lorentz contraction and time dilation," Am.J.Phys. **50**, 1156-1157 (1982)

[3] Asher Peres, "Relativistic telemetry," Am.J.Phys. **55,** 516-519 (1987)